\def \SAIT #1 #2 {{\em Mem.\ Soc.\ Astron.\ It.\/} {\bf #1}, #2}
\def \MESS #1 #2 {{\em The Messenger\/} {\bf #1}, #2}
\def \ASTRNACH #1 #2 {{\em Astron. Nach.\/} {\bf #1}, #2}
\def \AAP #1 #2 {{\em Astron. Astrophys.\/} {\bf #1}, #2}
\def \AAL #1 #2 {{\em Astron. Astrophys. Lett.\/} {\bf #1}, L#2}
\def \AAR #1 #2 {{\em Astron. Astrophys. Rev.\/} {\bf #1}, #2}
\def \AAS #1 #2 {{\em Astron. Astrophys. Suppl. Ser.\/} {\bf #1}, #2}
\def \AJ #1 #2 {{\em Astron. J.\/} {\bf #1}, #2}
\def \ANNREV #1 #2 {{\em Ann. Rev. Astron. Astrophys.\/} {\bf #1}, #2}
\def \APJ #1 #2 {{\em Astrophys. J.\/} {\bf #1}, #2}
\def \APJL #1 #2 {{\em Astrophys. J. Lett.\/} {\bf #1}, L#2}
\def \APJS #1 #2 {{\em Astrophys. J. Suppl.\/} {\bf #1}, #2}
\def \APSS #1 #2 {{\em Astrophys. Space Sci.\/} {\bf #1}, #2}
\def \ASR #1 #2 {{\em Adv. Space Res.\/} {\bf #1}, #2}
\def \BAIC #1 #2 {{\em Bull. Astron. Inst. Czechosl.\/} {\bf #1}, #2}
\def \JSQRT #1 #2 {{\em J. Quant. Spectrosc. Radiat. Transfer\/} {\bf #1}, #2}
\def \MN #1 #2 {{\em Mon. Not. R. Astr. Soc.\/} {\bf #1}, #2}
\def \MEM #1 #2 {{\em Mem. R. Astr. Soc.\/} {\bf #1}, #2}
\def \PLR #1 #2 {{\em Phys. Lett. Rev.\/} {\bf #1}, #2}
\def \PASJ #1 #2 {{\em Publ. Astron. Soc. Japan\/} {\bf #1}, #2}
\def \PASP #1 #2 {{\em Publ. Astr. Soc. Pacific\/} {\bf #1}, #2}
\def \NAT #1 #2 {{\em Nature\/} {\bf #1}, #2}
\def\gsimeq
{\hbox{\raise0.5ex\hbox{$>\lower1.06ex\hbox{$\kern-1.07em{\sim}$}$}}}
\def\lsimeq
{\hbox{\raise0.5ex\hbox{$<\lower1.06ex\hbox{$\kern-1.07em{\sim}$}$}}}
\def\pn{\par\noindent}

\def\ms{\medskip\pn}

\documentstyle[psfig]{memsait}
\begin{opening}
\title{DIFFUSE THERMAL EMISSION FROM VERY HOT GAS IN STARBURST GALAXIES: 
SPECTRAL RESULTS}
\vspace{-0.5truecm}
\author{
	M.~Cappi$^1$,
        M.~Persic$^2$,
        S.~Mariani$^3$,
        L.~Bassani$^1$,
        L.~Danese$^4$,
        A.J.~Dean$^5$,
        G.~Di~Cocco$^1$,
        A.~Franceschini$^6$,
        L.K.~Hunt$^7$,
        F.~Matteucci$^8$,
        E.~Palazzi$^3$,
        G.G.C.~Palumbo$^{1,3}$,
        Y.~Rephaeli$^9$,
        P.~Salucci$^4$, and 
        A.~Spizzichino$^1$}

\scriptsize
\institute{
$^1$ ITeSRE/CNR, via Gobetti 101, 40129 Bologna, Italy  
\\
$^2$ Trieste Astronomical Observatory, via G.B.Tiepolo 11, 
        34131 Trieste, Italy   
\\
$^3$Astronomy Dept., University of Bologna, via Zamboni 33, 
        40126 Bologna, Italy    
\\
$^4$SISSA/ISAS, via Beirut 2-4, 34013 Trieste, Italy        
\\
$^5$Physics Dept., Southampton University, Southampton SO9 5NH, UK 
\\
$^6$Astronomy Dept., University of Padova, vicolo dell'Osservatorio 5, 
        35122 Padova, Italy                        
\\
$^7$CAISMI/CNR, Largo E.Fermi 5, 50125 Firenze, Italy 
\\
$^8$Astronomy Dept., University of Trieste, via Besenghi, 
        34131 Trieste, Italy                              
\\
$^9$School of Physics and Astronomy, Tel Aviv University, Tel Aviv 69978,
        Israel\\}

\date{} 
\end{opening}

\normalsize
\begin{document}

\oddpagefooter{}{}{} 
\evenpagefooter{}{}{} 
\ms
\normalsize

\vspace{-1truecm}
\begin{abstract}

New $BeppoSAX$ observations of the nearby archetypical starburst 
galaxies (SBGs) NGC253 and M82 are presented. The main observational result 
is the unambiguous evidence that the hard (2-10 keV) component 
is (mostly) produced in both galaxies by thermal emission from a metal-poor 
($\sim$ 0.1--0.3 solar), hot (kT $\sim$ 6-- 9 keV) and extended (see 
companion paper: Cappi et al. 1998) plasma. 
Possible origins of this newly discovered component are briefly discussed. 
A remarkable similarity with the (Milky Way) Galactic Ridge's X-ray emission 
suggests, nevertheless, a common physical mechanism.

\end{abstract}

\vspace{-0.3cm}
\section{Main Interests}

The main interest of studying SBGs in the local universe 
is that it is believed that the starburst phase was very frequent 
during the early ($z \gsimeq 1$) evolution of galaxies (Madau et al. 1996). 
Thus understanding the properties of local 
SBGs means, presumably, understanding the properties of most galaxies of the 
early universe.  
Studies of SBGs can also help to shed light on key astrophysical issues like: i) the 
AGN/starburst connections, since both phenomena require concentration of 
material to the nuclear regions and are possibly linked to merging phenomena and/or 
presence of molecular bars (Maiolino et al. 1998), ii) the chemical enrichment 
of the intergalactic medium (IGM), because of the frequent presence  in SBGs 
of chemically-enriched SB-driven outflows (Heckman 1997) and iii) the structure, 
evolution and even 
formation of galaxies through the effect of the ``feedback'' provided by massive 
stars (Ikeuchi \& Norman 1991).

\vspace{-0.4cm}
\section{Current Observational Status}


Because of the short space available, we limit the following discussion to the 
2--10 keV energy band (see Fabbiano 1989 for a review on the optical and soft 
X-ray properties of SBGs).
In this energy range, the best-quality data available to date have been 
essentially limited to the two brightest SBGs NGC 253 and M82 (but see recent 
studies on lower-luminosity galaxies by Della Ceca et al. 1997). Data from 
the $Einstein$ IPC, $EXOSAT$ ME, $Ginga$, $BBXRT$ and $ASCA$ 
(Ptak et al. 1997 and ref. therein; Tsuru et al. 1997, Moran \& Lehnert 
1997) have clearly demonstrated the existence of a hard component well described by 
either a thermal model (kT $\sim$ 6--9 keV) or a power-law model ($\Gamma$ 
$\sim$ 1.8--2.0). The absence in the data of a significant Fe-K line emission 
has, however, always been puzzling and has induced several authors to propose 
explanations alternative to the thermal emission, i.e. the presence of a LLAGN
(Ptak et al. 1997, Tsuru et al. 1997) or inverse-Compton (IC) 
scattered emission from the interaction of IR photons with SNe-generated 
relativistic e$^{-}$ (Moran \& Lehnert 1997).

In the following, we present new $BeppoSAX$ results for these two 
SBGs which demonstrate that 
such hard component is most probably emitted by a thermal plasma. In this paper, 
spectral results showing the first evidences of Fe-K line emission (at $\sim$ 6.7 keV) 
and high-energy steepening expected in the case of thermal emission are presented for both 
NGC 253 and M82 (see Persic et al. 1998 for more details on the results for NGC 253).
In a companion paper (Cappi et al. 1998), we combine these spectral results 
to the (unprecedented) spatial information of $BeppoSAX$ between 2--10 keV.

\vspace{-0.4cm}
\section{BeppoSAX spectra of NGC 253 and M82}

$BeppoSAX$ observed NGC 253 on Nov.29--Dec.2, 1996 for $\sim$ 50, 113 and 50 Ksec
and M82 on Dec.06--07, 1997 for $\sim$ 29, 85 and 30 Ksec, for the LECS, MECS and PDS 
operating between 0.1--4 keV, 1.3--10 keV and 13--60 keV, respectively.
A Galactic column density of 1.28 $\times$ 10$^{20}$ cm$^{-2}$ for NGC 253 and 
4.27 $\times$ 10$^{20}$ cm$^{-2}$ for M82 were added in all the following spectral 
fits (Dickey \& Lockman 1990).
The spectral analysis was performed using version 10.00 of the XSPEC program.


\vspace{-0.5cm}
\begin{table}[hbt]
\begin{center}   
{\bf Table1:} Fe K line parameters \\
\begin{tabular}{cccc}
\hline
Source 	& kT$_{\rm brem}$ 	& Obs. Energy 	& EW \\
	& keV		& keV		& eV \\
\hline
NGC 253 & 7.40$^{+0.18}_{-0.71}$	& 6.69$^{+0.07}_{-0.07}$ & 329$^{+89}_{-109}$ \\
M82	& 9.27$^{+0.41}_{-0.51}$ & 6.65$^{+0.22}_{-0.19}$& 60$^{+35}_{-30}$ \\
\hline
\end{tabular}
\end{center}
\end{table}

\vspace{-0.5cm}
Several continuum models (e.g. a power-law, a bremsstrahlung, a Raymond-Smith, 
a multi-temperature Raymond-Smith, a Mewe-Kaastra-Liedahl model) 
and combination of them with and without absorption, 
with solar and variable abundances have been used to model the spectra (Fig. 1).
We have found that both spectra are best described in terms of a 
two-component thermal model (Fig. 2). At energies 
lower than $\sim$ 3 keV, emission lines (e.g. FeL, SiXV and SXV) characteristic 
of a soft thermal component are detected, in agreement with $ASCA$ results. 
At higher energies, both sources require 
the continuum emission to steepen (thus excluding a power-law description) and, most 
significantly, an Fe-K emission line at E $\sim$ 6.7 keV is detected in both 
objects (see Table 1 and Fig. 1). 
Interestingly, both hard components need to be absorbed in order not to 
over-produce the continuum at E $\lsimeq$1 keV (as also found with $ASCA$, Ptak et al. 1997). 

Best-fit results obtained by fitting both source spectra 
with a two-temperature Mewe-Kaastra-Liedahl model with variable abundances 
(``vmekal'' in XSPEC) are shown in Table 2 and Fig. 2.
Following Persic et al. (1998), in order to reduce the number of free parameters,
the abundances of He, C, and N were fixed at the solar value and the heavier
elements divided into two groups: Fe and Ni (most likely associated
to SNe I products), and the $\alpha$-elements O, Ne, Mg, Si, S, Ar and Ca
(most likely associated to SNe II products); elements in the same group
were constrained to have common abundances in solar units. 
It should be noted that although some extra contribution from a non-thermal 
(hard) component cannot be formaly excluded, we estimate that it should not 
contribute more than $\sim$ 20\% of the total 2--10 keV flux.

The detection of the Fe K line allows us to reliably determine the Fe abundances 
of the line-emitting matter: we find a value of $\sim$ 0.25 and $0.06$ solar for NGC 253 
and M82, respectively, consistent with the sub-solar values (based on upper-limits) 
predicted from previous hard X-ray observations (see \S 2). Although more insecure 
because of the current uncertainties of theoretical models in computing the 
Fe-L emissivity, similar Fe abundances are found for the soft component.
On the other side, $\alpha$-elements seem to have larger abundances ($\sim$ 1.7 solar 
for NGC 253 and $\sim$ 2.4 solar for M82).

\vspace{-0.5cm}
\begin{table}[htb]
\begin{center}
{\bf Table 2:} Two-components (vmekal) thermal models \\
\vspace{-0.2cm}
\begin{tabular}{cccccccc}
&&&& \\
\hline
Source & kT$_{soft}$ & Ab$_{soft}$ &  $N_{\rm H}^{\rm hard}$ & kT$_{hard}$ & 
Ab$_{hard}$ & $\chi^2/dof$ & log L \\
 & keV & $\alpha-$el. &  $\times 10^{22}$ & keV & $\alpha-$el. & & 2-10 keV\\
 & &  Fe, Ni & cm$^{-2}$ & & Fe, Ni & & erg/s \\ 
\hline 
NGC253 & $0.88_{-0.13}^{+0.13}$ & $1.69_{-0.72}^{+0.02}$ & $1.30_{-0.45}^{+0.63}$ & 
$5.60_{-0.60}^{+0.70}$ & $\leq 1.76$ & 244/237 & 1.6\\
& & $0.18_{-0.07}^{+0.29}$ & & & $0.27_{-0.08}^{+0.10}$ & &  \\
&&&&&& &  \\
 M82 & $0.70_{-0.05}^{+0.05}$ & $2.40_{-0.95}^{+1.31}$ & $0.51_{-0.17}^{+0.18}$ 
& $8.77_{-0.66}^{+0.53}$ & = Ab$_{soft}$ & 366/354 & 5.6 \\
& & $0.62_{-0.23}^{+0.40}$ & & & $0.07_{-0.03}^{+0.05}$ & & \\
\hline
\end{tabular}
\end{center}
\end{table}

\vspace{-0.9cm}
\section{On the origin of the hard, thermal, component}

On one side, our results for the soft X-ray component are in good 
agreement with previous X-ray results on SBGs (see \S 2), i.e. that it originates from 
a SB-driven galactic superwind (see Persic et al. 1998 for more details). 
On the other side, the origin of the hard thermal component is more puzzling. 
As demonstrated in Cappi et al. (1998), it is also extended possibly flowing 
out of the galaxy plane (as for the soft X-rays) in the case of NGC 253.
Among possible candidates, one could consider a collection of point sources 
(e.g. X-ray binaries, SNRs, clusters of protostars) or truly diffuse emission 
due to IC scattering of IR-optical 
photons from relativistic e$^{-}$ or, finally, a very hot 
($\sim$ 6--10 $\times$ 10$^7$ K) ISM phase. Alternatives to the hot ISM phase, however,
could hardly be reconciled with the observed spectrum because i) the average spectra of X-ray 
binaries would predict a power-law continuum and Fe K line at 6.4 keV, ii) the average 
spectra of SNRs would predict too softer (kT$\lsimeq$ 4 keV) spectra, iii) clusters 
of protostars have too low luminosities (L$_{\rm 2-10 keV}$ $\sim$ 10$^{32}$ erg s$^{-1}$)
and iv) IC emission would predict a power-law spectrum.
The most natural explanation seems therefore a hot thermal diffuse emission, i.e. 
a ``super super-wind'' most likely powered by the starburst.

It is interesting to note the striking analogy of the present results with 
the $ASCA$ observations of the Galactic Ridge (GR) (compare our Fig. 2 with 
Fig. 3 in Kaneda et al. 1997): also in this case, hot diffuse (and absorbed) 
emission is detected at a similar temperature and, if scaled for a 
$\sim$ 1/100 times lower Galactic star formation rate, with similar energetics. The only 
difference being in the somewhat larger ($\sim$ 0.8 solar) metal abundances 
deduced for the GR, these remarkable similarities strongly suggest a common 
physical origin.


\vspace{-0.4truecm} 
\section{Conclusions}

\vspace{-0.1cm}
The main conclusion from the above results is that the $BeppoSAX$ observations 
of the two archetypical SBGs NGC 253 and M82 have revealed for the first
time strong evidence of a hot, really hot, thermal diffuse plasma 
responsible for their hard (2--10 keV) 
emission, and that this discovery 
may have profound implications on AGN/starburst connections, 
on the chemical enrichment of the IGM and on the formation and evolution of galaxies.


\begin{figure}
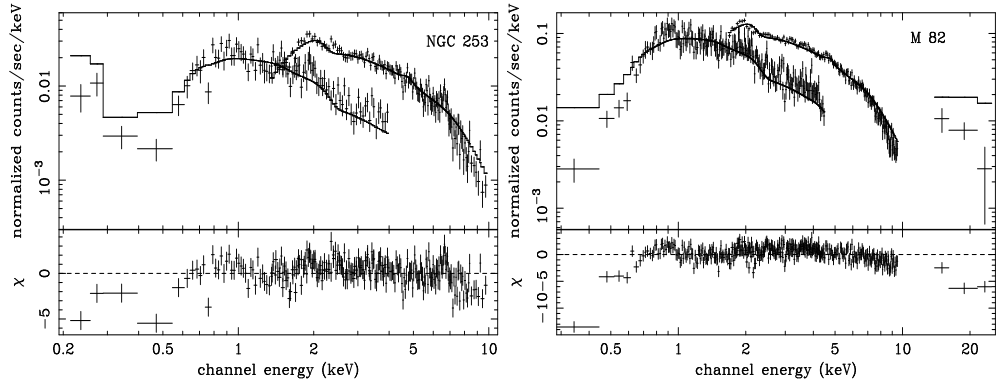

\vspace{-0.2cm}
\parbox{6.5truecm}
{\psfig{file=./cappi_fig1a.ps,width=6.5cm,height=5cm,angle=-90}}
\parbox{6.5truecm}
{\psfig{file=./cappi_fig1b.ps,width=6.5cm,height=5cm,angle=-90}}
\vspace{-0.2cm}
\caption[h]{Spectra of NGC 253 and M82 fitted with a single power-law model ($\Gamma$ 
$\sim$ 1.7) to illustrate the spectral complexities (emission lines and continuum curvature)}
\end{figure}
\begin{figure}
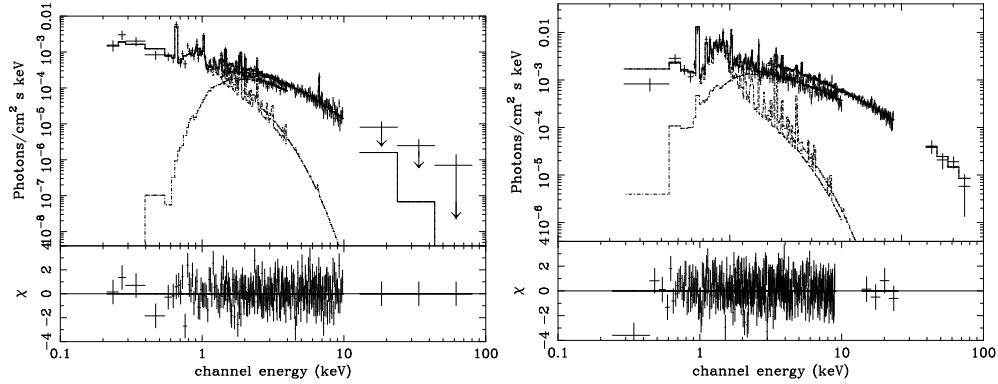

\vspace{-0.3cm}
\parbox{6.5truecm}
{\psfig{file=./cappi_fig2a.ps,width=6.5cm,height=5cm,angle=-90}}
\parbox{6.5truecm}
{\psfig{file=./cappi_fig2b.ps,width=6.5cm,height=5cm,angle=-90}}
\vspace{-0.2cm}
\caption[h]{Best-fit spectra of NGC 253 (left) and M 82 (right) with a 
2T Mekal model (Table 2).}
\end{figure}





\end{document}